\documentclass[sigconf]{acmart}
\usepackage{epigraph}
\usepackage{makecell}      % line breaks in headers
\usepackage{multirow}

\AtBeginDocument{%
  }

\setcopyright{acmlicensed}
\copyrightyear{2026}
\acmYear{2026}
\setcopyright{cc}
\setcctype{by}
\acmConference[CHI '26]{Proceedings of the 2026 CHI Conference on Human Factors in Computing Systems}{April 13--17, 2026}{Barcelona, Spain}
\acmBooktitle{Proceedings of the 2026 CHI Conference on Human Factors in Computing Systems (CHI '26), April 13--17, 2026, Barcelona, Spain}
\acmPrice{}
\acmDOI{10.1145/3772318.3790908}
\acmISBN{979-8-4007-2278-3/2026/04}

\begin{document}

\title[Envisioning the Future of K-12 GenAI Education from Global Teachers’ Perspectives]{Do Teachers Dream of GenAI Widening Educational (In)equality? Envisioning the Future of K-12 GenAI Education from Global Teachers’ Perspectives}

\author{Ruiwei Xiao}
\authornote{Both authors contributed equally to this research.}
\affiliation{
  \institution{Carnegie Mellon University}
  \city{Pittsburgh}
  \state{Pennsylvania}
  \country{USA}
}
\email{ruiweix@cs.cmu.edu}

\author{Qing Xiao}
\affiliation{
  \institution{Carnegie Mellon University}
  \city{Pittsburgh}
  \state{Pennsylvania}
  \country{USA}
}
\email{qingx@cs.cmu.edu}
\authornotemark[1]

\author{Xinying Hou}
\affiliation{
  \institution{University of Michigan}
  \city{Ann Arbor}
  \state{Michigan}
  \country{USA}
}
\email{xyhou@umich.edu}

\author{Phenyo Phemelo Moletsane}
\affiliation{
  \institution{Carnegie Mellon University}
  \city{Pittsburgh}
  \state{Pennsylvania}
  \country{USA}
}
\email{pmoletsa@andrew.cmu.edu}

\author{Hanqi Jane Li}
\affiliation{
  \institution{\mbox{University of California, San Diego}}
  \city{La Jolla}
  \state{California}
  \country{USA}
}
\email{jal221@ucsd.edu}

\author{Hong Shen}
\affiliation{
  \institution{Carnegie Mellon University}
  \city{Pittsburgh}
  \state{Pennsylvania}
  \country{USA}
}
\email{hongs@cs.cmu.edu}

\author{John Stamper}
\affiliation{
  \institution{Carnegie Mellon University}
  \city{Pittsburgh}
  \state{Pennsylvania}
  \country{USA}
}
\email{jstamper@cmu.edu}

\renewcommand{\shortauthors}{Xiao et al.}

\begin{abstract}
Generative artificial intelligence (GenAI) is rapidly entering K-12 classrooms worldwide, initiating urgent debates about its potential to either reduce or exacerbate educational inequalities. Drawing on interviews with 30 K-12 teachers across the United States, South Africa, and Taiwan, this study examines how teachers navigate this GenAI tension around educational equalities. We found teachers actively framed GenAI education as an equality-oriented practice: they used it to alleviate pre-existing inequalities while simultaneously working to prevent new inequalities from emerging. Despite these efforts, teachers confronted persistent systemic barriers, i.e., unequal infrastructure, insufficient professional training, and restrictive social norms, that individual initiative alone could not overcome. Teachers thus articulated normative visions for more inclusive GenAI education. By centering teachers’ practices, constraints, and future envisions, this study contributes a global account of how GenAI education is being integrated into K-12 contexts and highlights what is required to make its adoption genuinely equal.
\end{abstract}



\maketitle
\section{Introduction}

\begin{quote}
\itshape
``It was the best of times, it was the worst of times, it was the age of wisdom, 
it was the age of foolishness, it was the epoch of belief, it was the epoch of incredulity, 
it was the season of Light, it was the season of Darkness, it was the spring of hope, 
it was the winter of despair, we had everything before us, we had nothing before us, 
we were all going direct to Heaven, we were all going direct the other way.''

\hfill --- Charles Dickens, \textit{A Tale of Two Cities}
\end{quote}

\vspace{8mm}

Today, generative artificial intelligence (GenAI) is rapidly entering K–12 classrooms across the globe \cite{mintz2023artificial,mathur2025ai,xiao2025ai}. Whether teachers welcome it or not, students are already engaging with GenAI tools in and beyond the classroom \cite{adisa2025middle}, for instance, using ChatGPT to draft essays \cite{levine2025students}, solve math problems \cite{walkington2025middle}, or generate summaries of readings \cite{steele2023gpt}. This rapid spread makes GenAI not only a technological trend but also an urgent equality issue at the societal level. On one hand, it has the potential to democratize learning opportunities, providing personalized tutoring and cognitive support that could help close long-standing gaps in access and achievement \cite{tafazoli2024exploring,james2024levelling}. On the other hand, it risks exacerbating divides by disproportionately benefiting students with reliable infrastructure, digital literacy, and supportive school resources, while leaving behind those in under-resourced contexts \cite{hendawy2024intensified,murgia2024uninvited}.

Much like Dickens’s portrayal of an era marked by both promise and peril \cite{dickens2007tale}, GenAI in education holds the potential to be a “season of Light,” lowering entry barriers and empowering historically underserved learners, or a “season of Darkness,” where algorithmic bias, plagiarism, and overreliance risk deepening existing divides \cite{selwyn2024constructive}. From a sociotechnical perspective, technologies do not produce outcomes on their own, as they acquire meaning through human negotiation, institutional norms, and infrastructural conditions \cite{orlikowski1992duality,star1999ethnography,ackerman2000intellectual,li2026moon}. K–12 teachers, who often function as everyday mediators of societal issues within classrooms \cite{kaka2023social}, therefore stand at the front line of this tension: for them, GenAI’s double-edged nature is not speculative, but a daily reality requiring judgment, interpretation, and pedagogical action. K-12 teachers must navigate how GenAI simultaneously opens new educational opportunities and deepens existing inequities. Yet despite its urgency, most existing work continues to examine adoption factors \cite{li2024explanatory,kong2024examining,liu2025adopting} or broad risks in the use of GenAI tools in education \cite{hoernig2024generative,newton2025vulnerable,sallai2024approach,harvey2025don}. What remains underexplored is how teachers themselves, those directly mediating young students’ interaction with GenAI, grapple with this tension in their everyday practices and attempt to \textit{teach} GenAI in ways that promote, rather than undermine, educational equality. 

To address these gaps, we focus on how K–12 teachers currently teach their students about GenAI, with particular attention to the global commonalities of the tensions it creates regarding educational equality. GenAI is not reshaping K–12 education in any single region; instead, it is transforming educational opportunities and inequalities worldwide. These dynamics call for a global perspective to study not only how GenAI is currently being taught in K–12 settings worldwide, but also how it should be taught in the future to foster greater educational equality. 

We selected three regions: the United States, South Africa, and Taiwan, as illustrative cases, focusing on their K–12 education systems. In the U.S., even within a highly resourced system, internal resource inequalities in K–12 schooling can perpetuate unequal opportunities \cite{hai2025,magicschool}. In South Africa, deep systemic inequality due to colonization, linguistic diversity (e.g., 12 official languages), and infrastructural scarcity strongly shape how K–12 classrooms engage with new technologies \cite{spaull2019south,statssaQLFS2025Q2,southafrica_gateway2025,doe2006caps,doe2015catsyllabus}. In Taiwan, strong curricular mandates and government support extend into K–12 schools to advance AI literacy, yet linguistic (e.g., code-switching between Traditional and Simplified Chinese) and cultural particularities (e.g., indigenous cultural representation and local identity) complicate equitable adoption \cite{lyu2025characterizing,ndc2025twdigital}. We provide further details about these contexts in \autoref{context}. Examining these contexts together foregrounds how K-12 GenAI education carries both the promise of reducing inequality and the risk of reinforcing it, depending on how teachers and broader society respond.

Specifically, in this paper, we answer these questions across global, structurally diverse educational systems:  
\begin{itemize}
    \item \textbf{RQ1:} How do K–12 teachers integrate GenAI education into their classrooms to promote educational equality?  
    \item \textbf{RQ2:} What structural challenges beyond teachers’ control do K–12 teachers encounter when teaching GenAI to promote educational equality?  
    \item \textbf{RQ3:} Beyond their own classroom efforts, what kinds of support and future directions do teachers envision from broader stakeholders (e.g., schools, communities, policymakers, and technology companies) to make K–12 GenAI education more equal?  
\end{itemize}

Our study draws on in-depth interviews with 30 K–12 AI education teachers, 10 from each of the United States, South Africa, and Taiwan, who are actively guiding students in using GenAI and reflecting on its role in promoting or undermining educational equality. Some participants worked in schools with formal information technology (IT) or artificial intelligence (AI) curricula, while others introduced GenAI in contexts where such instruction was not institutionalized. Across these settings, teachers consistently framed GenAI not merely as a new technological tool, but as a pedagogical site for navigating inequality. They described their GenAI teaching practices as attempts to use GenAI to alleviate pre-existing educational inequities while simultaneously working to prevent the emergence of new inequalities. 

In this study, we define \textit{GenAI education} as instruction aimed at building learners’ capacity to properly interact with GenAI tools, drawing on the widely accepted definition or scope of AI Education in HCI and Learning Sciences communities \cite{tadimalla2025ai,long2020ai,lin2021engaging} but with a narrowed focus on generative AI. In K–12 settings, this includes teaching students foundational knowledge about what GenAI is and how it works, creating opportunities for hands-on interaction with GenAI systems, and guiding learners to embed GenAI into regular subject learning (e.g., writing, science, or social studies), even in the absence of a formal AI course. It also involves guiding students to critically reflect on the benefits, risks, and responsible use of GenAI. By contrast, our scope does not include the closely related but distinct concept of \textit{GenAI in education}, which refers to the use of GenAI technology to enhance classroom/education experiences \cite{tadimalla2025ai}. Instead, we focus specifically on teachers’ efforts to \textit{teach} students about GenAI as part of their K–12 learning experience. \label{def_genai_ed}

Our findings are organized around three interrelated themes. Firstly, we explored current equality-oriented teaching practices in GenAI education. Teachers across all three regions described concrete strategies for using GenAI education to mitigate pre-existing inequalities, such as gaps in technology awareness, access to resources, digital literacy, and opportunities for marginalized learners, while also embedding practices to prevent GenAI from creating new disparities, including unfair shortcuts in college admission, misuse for bullying, cultural erasure of local, low-resourced languages, and overreliance on GenAI (\textit{\textbf{RQ1}}). Secondly, we found that despite these equality-oriented practices, teachers encountered structural challenges that individual commitment alone could not overcome. Infrastructural divides, insufficient training and curricular guidance, and restrictive social norms repeatedly constrained their ability to integrate GenAI equitably across classrooms (\textit{\textbf{RQ2}}). Thirdly, looking ahead, teachers articulated practical directions for building more inclusive GenAI education with additional support from other stakeholders. Their visions spanned school-level innovation centers and outreach to redistribute resources, company-level design changes (e.g., cultural localization of AI tools), and government-level initiatives (e.g., AI literacy for parents and communities). These forward-looking perspectives framed GenAI education not only as a classroom practice but also as a shared societal responsibility to improve educational equalities (\textit{\textbf{RQ3}}).

This study makes three key contributions to future HCI and education research:

\begin{itemize}
    \item First, we extend prior AI education frameworks by documenting how global teachers are already embedding GenAI into their classrooms, connecting their pedagogical strategies to long-standing patterns of educational (in)equality.
    \item Second, we highlight how infrastructural inequalities, gaps in training, and AI-related stigma constrain the reach of individual teacher efforts, showing that systemic conditions shape how equitably students can benefit from GenAI.
    \item Third, we bridge existing AI education efforts with future HCI agendas by surfacing teachers’ forward-looking proposals for GenAI integration across stakeholder levels. These insights not only translate into actionable implications for schools, companies, and governments, but also extend prior HCI frameworks by identifying unresolved sociotechnical challenges and outlining where future methodological and design innovations are needed to advance equitable GenAI education.
\end{itemize}

\section{Related Work}
Our study is built on two threads of literature: how K–12 teachers teach students with and about AI (see \autoref{lr1}), and how long-standing educational inequalities are reshaped in the GenAI era (see \autoref{lr2}).

\subsection{Teaching with and about AI in K–12 Education} \label{lr1}

Ongoing discussions continue around the integration of GenAI into K–12 classrooms \cite{zhang2024systematic,whalen2025k,laak2024generative,hidalgo2025generative}. While most work highlighted AI's potential on personalized learning and high-quality instructional materials \cite{zhang2024systematic, mintz2023artificial}, many also acknowledged its risks, such as impacts on academic integrity \cite{zhang2024systematic}, algorithmic bias, data privacy, and questioned whether the benefits outweigh them \cite{cheah2025integrating}. For example, \citeauthor{Han2024CHI} found that teachers, parents, and students liked LLMs for adaptable materials, idea generation, and timely feedback. But they required adult oversight, appropriate control over AI roles, and customization \cite{Han2024CHI}. Within this context, HCI researchers have examined the responsible use of GenAI in K-12 education. For instance, \citeauthor{zhu2025bridging} has investigated how a GenAI-powered math story creator can benefit math instruction and learning \cite{zhu2025bridging}. In addition, \citeauthor{kim2024app} investigated the effectiveness of introducing a GenAI chatbot into a mobile app development course \cite{kim2024app}. 

This broader availability of GenAI tools is increasingly shaping the digital and educational experiences of K-12 students \cite{grover2024teaching}. As educators established multi-level concerns about GenAI tools in K-12 contexts, including academic dishonesty and student social development restriction \cite{harvey2025don}, many recent explorations emphasize GenAI Education in the K-12 context, such as teaching students the concepts, uses, and responsible practices of GenAI \cite{su2022meta, grover2024teaching, lee2024systematic}. One line of studies focused on integrating AI education into the existing K-12 core subject curriculum \cite{Lin2021CHI,lee2022preparing, park2023integrating}, such as science \cite{park2023integrating, wan2020smileycluster} and math domains \cite{wang2019ai}. Another line of research has focused on offering more specialized AI education \cite{gibellini2023ai,Cao2025CHI}. For instance, \citeauthor{gibellini2023ai} worked with K-12 teachers to explore how to include an AI curriculum in middle school inclusively \cite{gibellini2023ai}. Similarly, \citeauthor{lin2021engaging} co-designed an AI curriculum with 15 K-12 teachers \cite{lin2021engaging}. In addition, \citeauthor{Cao2025CHI} co-designed with students to embed the unique history and community culture into AI literacy education \cite{Cao2025CHI}. Besides explorations of curriculum and education design, researchers also investigated a broad scope of activities and technologies \cite{jia2025technologies} to help K-12 students learn about AI, such as shorter-term workshops \cite{klemettila2025s}, unplugged activities \cite{darabipourshiraz2025introducing}, and games \cite{minecraft2025}. 

Most work explores GenAI integration from a technology perspective, or AI curricular content from a learning sciences perspective, but less is known about why and how non-AI-specialist teachers teach GenAI in everyday practice and what are the cultural and sociological challenges from a sociotechnical perspective. This gap is critical for HCI/CSCW, as technologies gain meaning through situated negotiation rather than deployment alone \cite{long2020ai,holstein2019designing,orlikowski1992duality}, and understanding these societal, empirical factors helps improve GenAI-related products, design relatable GenAI learning experiences, and supplement existing human-AI interaction design guidelines \cite{amershi2019guidelines}. As teachers are active mediators who shape students' interaction with, and participation in, everyday AI experiences and futures with AI \cite{holstein2019designing,harvey2025don}, by adopting teachers’ lens, this work addresses how GenAI is taught as a sociotechnical practice beyond tool adoption or knowledge transmission \cite{ackerman2000intellectual,star1999ethnography}, contributing a human-centered view to human–GenAI interaction.

\subsection{Tracing Educational (In)equalities: From Traditional Schooling to the GenAI Era} \label{lr2}

Educational inequality has long been a central concern in K-12 education. Before the emergence of GenAI, scholars already documented persistent achievement and attainment gaps rooted in socioeconomic background, family resources, school quality, and peer environments. Cross-national studies show how higher income inequality is associated with lower social mobility, with educational inequality functioning as a key mechanism in this process \cite{Blanden2022HES,Corak2013JEP,hassler2007inequality}. Parental resources and peer effects reinforce these learning gaps \cite{BeckerTomes1979JPE,BisinVerdier2001JET,EppleRomano2011HSE,li2026moon}, and global assessments like PISA continue to highlight enduring socioeconomic divides in learning outcomes \cite{OECD2016PISA,Blanden2022HES}.

With the rise of GenAI, these dynamics take on new forms \cite{capraro2024impact,Kunkeler2022UKICER}. On one hand, studies show potential for AI to enhance engagement, support learners with disabilities, and reduce gaps when paired with ethical frameworks and accessible design \cite{Rasheed2025InequalityAI,Kunjumuhammed2024AIineqHEI,XIA2025100455}. For example, AI-powered intelligent tutoring system can help students achieve better academic performance \cite{akgun2022artificial}. \citeauthor{XIA2025100455} reported that GenAI had significant positive effects on students’ motivation and engagement across diverse educational settings \cite{XIA2025100455}. On the other hand, critical work emphasizes that without adequate infrastructure, localized tools, affordability, registers IT teachers and IT courses offered, GenAI risks amplifying rather than alleviating existing inequalities \cite{Ahmed2024DigitalDivideAI,Hussein2025DigitalInequalities}. For example, \citeauthor{joshi2020state} classified the world’s languages by resource availability: only 7 (0.28\%, spoken by 2.5 billion people) are high-resource, while the remaining 99.7\% spoken by 5 billion people are mid- or low-resource \cite{joshi2020state}. Consequently, language models perform significantly worse on these underrepresented languages \cite{lyu2025characterizing}. Such uneven data representation also raises ethical concerns in education: in their ethics of AI in education framework, \citeauthor{holmes2022ethics} report that 17 AI in Education (AIED) community’s leading researchers warned of potential disadvantages for marginalized groups due to biased data \cite{holmes2022ethics}. Inequalities also emerge from educational capacity: in some regions, IT is not compulsory, whereas in Taiwan, nearly 2,000 certified IT teachers serve middle schools and over 2,300 serve high schools \cite{moe2023itteachers}.

Prior HCI/CSCW works have shown that educational technologies do not translate into equitable outcomes automatically, but must be interpreted, negotiated, and appropriated by human actors within institutional and cultural constraints \cite{cao2025ai,orlikowski1992duality,ackerman2000intellectual}. Beyond empirical studies, policy and industry analyses such as \citet{luckin2016intelligence}’s report in 2016, highlighting that AI in education will only benefit learners equitably when teachers are supported as orchestrators of AI use \cite{luckin2016intelligence}. Extending this perspective to nowadays, GenAI-in-school contexts, we aim to understand how teachers mobilize GenAI education to promote equality, and what organizational and societal conditions enable or hinder such work, connecting GenAI education to broader CSCW debates on infrastructure, labor, and mediation, and foregrounds teachers’ work as a gateway to designing equitable futures with AI.

\section{Method}
We conducted 30 in-depth, semi-structured interviews with K–12 teachers who were actively engaged in teaching GenAI in their classrooms and who expressed an interest in using or critically reflecting on GenAI in relation to issues of educational (in)equality. The sample included 10 teachers, each from the United States, South Africa, and Taiwan. Each interview lasted between 60 and 90 minutes, was conducted remotely via Zoom, and participants received USD 20 per hour as compensation. Our focus was on how teachers perceive the double-edged nature of GenAI, both as a tool for expanding inclusive opportunities and as a risk for deepening divides, and how these perceptions shape the way they teach GenAI to their students.

\subsection{Context: AI Education in the United States, South Africa, and Taiwan} \label{context}
We focus on the United States, South Africa, and Taiwan to examine both the shared challenges and context-specific dynamics of K-12 GenAI education. By studying these diverse contexts together, we foreground how (in)equality issues in K-12 GenAI education are marked by both universal tensions and situated differences.
% A Table here
\subsubsection{United States.}
The U.S. leads in AI development and educational technology, with English as a high-resource language and most widely used AI tools (e.g., ChatGPT, Khanmigo) developed domestically \cite{hai2025,magicschool}. Its decentralized K–12 system means states and districts determine curricula and technology adoption \cite{usdoe2020}, though federal initiatives like the 2025 Executive Order on AI education provide guidance \cite{whitehouse2025aiYouth}. Most AI concepts are taught within computer science electives, with broad device access achieved since the pandemic \cite{whitehouse2025aiYouth,edweek2022devices}. By March 2021, 90\% of districts reported a device for every middle/high school student and 84\% for elementary students \cite{edweek2022devices}.

\subsubsection{South Africa.}

South Africa’s AI education is shaped by stark socio-economic inequality, youth unemployment, and linguistic diversity \cite{spaull2019south,statssaQLFS2025Q2}. Limited infrastructure and uneven school resources perpetuate inequalities in the distribution of national educational resources, especially between urban schools and township or rural schools \cite{sibuyi2024investigation}. To promote educational equality in the GenAI era and prepare the next generation for digital futures, government initiatives such as the Basic Education Employment Initiative (BEEI) have been introduced to improve digital readiness \cite{southafrica_gateway2025}. Teachers face the challenge of adapting AI use across 12 official languages while addressing unequal access to devices and training \cite{southafrica_gateway2025}. Although demand for AI is high \cite{mckinsey2025AfricaGenAI}, local languages remain low-resource, and most tools in use are U.S.-developed \cite{businesstech2024}, which can limit their effectiveness in South Africa due to mismatches with local languages, curricula, culture and classroom contexts. Domestically, there are startups specialize in AI for education, but their growth is constrained by weak infrastructure, skills shortages, and regulatory uncertainty \cite{edtechsa2025}. In curriculum, IT is not compulsory: Grades R–9 include only basic digital literacy \cite{doe2006caps}, while Grades 10–12 offer two electives, Computer Applications Technology (CAT) and Information Technology (IT), with access highly unequal between elite and under-resourced schools \cite{doe2006caps,doe2015catsyllabus}.

\subsubsection{Taiwan.}
Taiwan combines advanced high-tech development with cultural and linguistic particularities. Under the centralized “108” Curriculum Guidelines, IT (including AI literacy) is mandatory from middle school, supported by nationwide teacher-training platforms like the Adaptive Learning Platform (ADL) \cite{moeadl,Lee2024studentsView108}. The government funds devices and resources for rural and indigenous schools to reduce disparities \cite{ndc2025twdigital}. However, differences between Traditional Chinese and Simplified Chinese corpora pose challenges for localizing large language models, given the larger scale and dominance of Simplified corpora compared to the smaller, Taiwan-focused Traditional corpora \cite{lyu2025characterizing}.

\subsection{Participants: Global K–12 GenAI Education Teachers}

We recruited participants based on three main criteria: (1) they were active K–12 teachers with direct classroom responsibilities; (2) they were teaching GenAI, either through dedicated AI- or IT-related courses or by incorporating GenAI into their own subject areas (e.g., English, social science, or STEM); and (3) they demonstrated a particular interest in the relationship between GenAI and educational (in)equality. Teachers in South Africa and the U.S. were randomly sampled through Prolific, while teachers in Taiwan were recruited via mailing lists due to the smaller eligible pool (n<10) on Prolific. A pre-screening survey was used to ensure alignment with the three criteria.

\begin{table*}[ht]
\centering
\caption{Participant Demographics.}
\label{tab:participants}
\footnotesize
\renewcommand{\arraystretch}{1.25} 
\resizebox{\linewidth}{!}{%
\begin{tabular}{l l l l l l l l}
\toprule
\textbf{ID} & \textbf{Location} & \textbf{Gender} & \textbf{Years of Experience} & \textbf{School Type} & \textbf{K--12 Level} & \textbf{Teaching Subject} & \textbf{AI-Related Course} \\
\hline
U1  & US, Florida        & Male   & 9   & Public    & High School   & Social Science & N/A \\
U2  & US, Wisconsin      & Female & 14  & Public            & High School   & English & N/A \\
U3  & US, New Jersey     & Female & 26  & Public            & High School   & Statistics, CS, Math & CS course \\
U4  & US, Texas          & Male   & 11  & Public            & High School   & English, Creative Writing & N/A \\
U5  & US, Oregon         & Female & 5   & Public            & Elementary School   & Special Education & IT course \\
U6  & US, Nebraska       & Male   & 27  & Public            & High School   & Social Studies, History, Civics & N/A \\
U7  & US, Midwest        & Female & 7   & Public            & High School   & English & N/A \\
U8  & US, Mississippi    & Female & 18  & Public            & High School   & English, Social Studies, Economics & N/A \\
U9  & US, Florida        & Female & 12  & Private & Elementary School   & Science, Social Studies & N/A \\
U10 & US, Michigan       & Female & 27  & Public            & High School   & English, History & N/A \\
\hline
T1  & Taiwan, Hsinchu    & Female & <1  & Public            & Middle School   & Information Technology & IT course \\
T2  & Taiwan, Hualien    & Male   & 29  & Public            & Elementary School   & Information Technology, Arts & IT course \\
T3  & Taiwan, Taipei     & Female & 1.5 & Public            & Kindergarten  & Life Skills & N/A \\
T4  & Taiwan, Hsinchu    & Female & 7   & Public            & Elementary School   & Chinese, Math, Social skills & IT course \\
T5  & Taiwan, Taipei     & Female & 22  & Public            & High School   & Information Technology & IT course \\
T6  & Taiwan, Hsinchu    & Female & 15  & Public            & Middle School   & Information Technology & IT course \\
T7  & Taiwan, Taipei     & Male   & 33  & Public            & High School   & Information Technology & IT course \\
T8  & Taiwan, Banqiao    & Male   & 7   & Private & High School & STEM & CS course \\
T9  & Taiwan, Hsinchu    & Female & <1  & Public            & Elementary School   & Special Education & IT course \\
T10 & Taiwan, Hsinchu     & Female & 2   & Public            & Kindergarten    & Life Skills & N/A \\
\hline
S1  & South Africa, KwaZulu-Natal & Male   & 3   & Public  & High School   & English, Sepedi (Local Language) & N/A \\
S2  & South Africa, Gauteng     & Female & 5   & Public  & High School   & Math, English & N/A \\
S3  & South Africa, Limpopo     & Female & 15  & Public  & High School   & English, Afrikaans (Local Language) & N/A \\
S4  & South Africa, Pumalanga   & Female & <1  & Public  & Elementary School, High School & Math, Science, English, PE & N/A \\
S5  & South Africa, KwaZulu-Natal & Female & 5 & Public  & High School   & Business, Life Orientation & N/A \\
S6  & South Africa, KwaZulu-Natal & Female & 2 & Public  & High School   & English, Economics, Tourism & IT course \\
S7  & South Africa, Mpumalanga  & Male   & 15  & Private & High School   & Math, Life Sciences & N/A \\
S8  & South Africa, Johannesburg & Female & 8  & Private     & High School   & Science, Chemistry & CS course \\
S9  & South Africa, KwaZulu-Natal & Female & 4 & Public  & Preschool, Elementary School & Math, English, Life Skills & IT course \\
S10 & South Africa, Free State  & Male   & 2   & Public  & Elementary School   & English, Social Sciences, Life Skills & N/A \\
\bottomrule
\end{tabular}%
}
\end{table*}

In total, we interviewed 30 teachers, with 10 each from the United States, South Africa, and Taiwan. Table~\ref{tab:participants} presents the demographic and professional characteristics of our participants. Across the three regions, our sample included a representative mix of genders and school types. In the United States, we interviewed 10 teachers (3 male, 7 female), the majority of whom were from public schools, with one participant from a private elementary school. In Taiwan, the 10 participants included 3 male and 7 female teachers, almost all of whom were from public schools, with one participant from a private high school. In South Africa, we also interviewed 10 teachers (3 male, 7 female); most were based in public schools, with two participants from private high schools.
Regarding their GenAI teaching practices, some, particularly in Taiwan, taught formal IT or computer science courses that included AI-related content as part of their curriculum. Others, especially in the U.S., South Africa, and in non-technical subjects, integrated GenAI into existing classes, using it to support essay writing, creative projects, science assignments, or critical discussions about digital literacy and bias. Teaching experience ranged from less than one year to over three decades.

\subsection{Semi-Structured Interviews}
This study was approved by the IRB (Institutional Review Board). We focus on how K-12 teachers teach GenAI to their students in our interviews. Each interview lasted between 60 and 90 minutes and was conducted remotely via Zoom. To accommodate participants and ensure they could express themselves comfortably, interviews were conducted in the primary teaching language of each teacher. All interviews were audio-recorded, transcribed, and, when necessary, translated into English for analysis. While all participants were asked the same core set of questions, follow-up prompts were adapted to each context, encouraging teachers to provide concrete classroom examples, reflections on their motivations, and their perceptions of GenAI’s double-edged role as both an equalizer and a potential amplifier of inequality when teaching their students.

We began each interview with questions about teachers’ professional trajectories and school contexts. Participants introduced themselves by describing their years of teaching experience, subject areas, and current teaching roles. We asked about grade levels and student populations, school type, and the socio-economic backgrounds of their students. Teachers also reflected on the IT infrastructure available in their schools and how their personal background shaped their understanding of GenAI education. 

We then asked teachers to describe how they introduced GenAI to their students in daily teaching. We invited them to share classroom activities, lesson plans, or assignments where GenAI played a role, and to reflect on whether these GenAI education practices reduced or reproduced pre-existing inequalities. Teachers were also asked about their pedagogical motivations, for instance, whether they saw GenAI as a way to expand access to resources for disadvantaged learners. The following section focused on the broader conditions (e.g., school environment, local policies, technology infrastructures) that shape GenAI education beyond individual teachers’ efforts. We explored infrastructural issues, including uneven access to devices and internet connectivity, as well as gaps in professional training, unclear curricula, and restrictive school and district policies. Teachers also discussed cultural and institutional norms that influenced their GenAI education practice, including parental attitudes, school climate, policy directives, and the role of EdTech companies. The final part asked teachers to articulate their aspirations and proposals for making GenAI education more inclusive. We encouraged them to imagine what kinds of tools, policies, or support systems would be most helpful for their future GenAI education. 

We also tailored follow-up questions to reflect region-specific conditions. In the United States, we asked teachers how GenAI intersected with decentralized governance and state-level curriculum standards. In South Africa, we investigated how infrastructural challenges (e.g., intermittent electricity, uneven internet access), linguistic diversity, and persistent socioeconomic inequality influenced their GenAI teaching practices. In Taiwan, we explored how national curriculum mandates (e.g., the 108 Curriculum Guidelines), and the distinction between Traditional and Simplified Chinese influenced teachers’ classroom practices. Alongside these region-specific probes, we also incorporated culturally oriented follow-up questions. Teachers were encouraged to reflect on how local educational philosophies, community expectations, and broader cultural norms shaped their motivations and strategies for teaching GenAI. 

\subsection{Data Analysis}

We analyzed the interview data using a reflexive thematic analysis approach \cite{braun2006using,braun2019reflecting}. First, two members of the research team independently reviewed the transcripts and developed initial codes that captured recurring practices, challenges, and aspirations described by participants. We then iteratively refined the coding framework, combining inductive codes that emerged from the data with deductive codes derived from our research questions and prior work on AI education and educational inequality. To ensure reliability and depth, the team engaged in multiple rounds of coding comparison and discussion, resolving discrepancies through deliberation rather than statistical measures of inter-rater agreement, in line with traditional practices in qualitative HCI research \cite{xiao2025might}. Themes were developed by clustering related codes, with particular attention to how teachers’ narratives reflected both global commonalities and context-specific conditions. Throughout the process, we wrote analytic memos to trace connections across regions and to foreground teachers’ own framings of how GenAI can either mitigate or exacerbate inequalities in K–12 education.  

In analyzing data across the three regions, we adopted a comparative yet integrative approach. We first examined each region separately to ensure that locally specific practices and challenges were not overshadowed by broader trends. This allowed us to capture, for example, how Taiwan’s centralized curriculum structure shaped teachers’ opportunities to formalize GenAI education, or how South Africa’s infrastructural and linguistic constraints created distinctive barriers. We then moved across cases to identify thematic overlaps, such as shared concerns about equal access, digital literacy gaps, or the risks of plagiarism and bias. By moving iteratively between within-region and cross-region analysis, we were able to highlight both the global commonalities of how teachers engage with GenAI education and the situated differences that condition its use. This dual focus ensured that our findings speak to universal challenges in GenAI education while remaining grounded in the particularities of diverse educational systems. The emerging themes of our analysis match our finding structure (see \autoref{RQ1}, \autoref{RQ2}, \autoref{RQ3}).

\subsection{Positionality}

Before beginning this study, we engaged deeply with the sociocultural and educational contexts of the three regions. Our preparation included reviewing local education policies, government reports, and media coverage related to AI and schooling, as well as drawing on prior research on issues of inequality in each setting. These materials helped us to conduct interviews that were sensitive to local conditions and to interpret teachers’ accounts in light of broader structural and cultural dynamics. 

Our international author team includes at least one member from each cultural region who was born there and has conducted educational research within that setting for a minimum of two years. This collaborative structure enabled us to triangulate perspectives, verify interpretations against local knowledge, and remain mindful of how our own academic and cultural backgrounds influenced the questions we asked and the themes we emphasized. 

\subsection{Ethical and Reflexive Considerations}

In addition to formal IRB approval, this study required ongoing ethical reflection due to the sensitive nature of K--12 education and the involvement of teachers working with legally and socially protected populations. While our interviews focused on teachers’ pedagogical practices rather than collecting data from students directly, teachers’ accounts frequently implicated issues of student data, surveillance, and emotional vulnerability in classroom uses of GenAI. We therefore approached data collection and analysis with heightened attention to privacy, power, and responsibility.

First, data privacy emerged as an implicit but significant ethical concern. Teachers often described using commercial GenAI platforms whose data practices were opaque to them, particularly with respect to how student prompts, outputs, or interaction logs might be stored or reused. Although our study did not examine platform data flows directly, we recognize that teachers’ uncertainty around data governance shaped both their teaching practices and their ethical hesitation. In reporting these accounts, we deliberately avoided collecting or reproducing any identifiable student data and treated teachers’ descriptions of classroom incidents at an abstract level to prevent indirect disclosure of sensitive information.

Second, our cross-national design foregrounded power asymmetries that operate differently across policy regimes. Teachers in relatively centralized systems (e.g., Taiwan) and decentralized systems (e.g., the United States) described distinct forms of constraint, yet across contexts, they shared a common position of being held responsible for ethical AI use without having commensurate authority over infrastructural decisions, platform selection, or data policies. As researchers, we are attentive to how documenting teachers’ struggles with GenAI risks normalizing this unequal distribution of responsibility. To mitigate this, our analysis consistently situates teachers’ practices within broader institutional and policy contexts rather than framing ethical challenges as individual shortcomings.

Third, we reflect on our positionality as researchers studying ethically contested technologies. While teachers often asked normative questions about what constitutes “responsible” GenAI use in classrooms, our role was not to prescribe correct practices but to surface the tensions and constraints that shape teachers’ decision-making. 

Together, these ethical and reflexive considerations reinforce one of our central claims: that equality-oriented GenAI education is shaped not only by pedagogical intentions, but also by asymmetric power relations and unresolved governance questions surrounding data, accountability, and control. Making these dynamics explicit strengthens the interpretive validity of our findings and underscores the need for ethical responsibility to be addressed at systemic, rather than solely individual, levels.

\section{Findings: Equality-Oriented Teaching Practices in K–12 GenAI Education (RQ1)} \label{RQ1}

\begin{table*}[t]
\centering
\footnotesize
\caption{Roles of GenAI education in alleviating existing educational inequalities and preventing the creation of new ones through teaching practices (RQ1)}
\setlength{\tabcolsep}{8pt}
\renewcommand{\arraystretch}{1.2}
\begin{tabular}{p{0.19\linewidth} p{0.31\linewidth} p{0.4\linewidth}}
\toprule
\textbf{GenAI Education Role} & \textbf{Education Inequality} & \textbf{Practices to Alleviate Inequality} \\
\midrule
\multirow{4}{=}{\textbf{Using GenAI Education to Alleviate Pre-Existing Educational Inequalities}}
& Consciousness inequality 
& Use GenAI education to bridge gaps in awareness and understanding of technology. \\

& Resource inequality 
& Use GenAI education to compensate for shortages in learning materials. \\

& Digital literacy inequality 
& Leverage GenAI accessibility to broaden computing participation. \\

& Lack of personalized learning for marginalized groups 
& Open opportunities for students marginalized by traditional educational structures. \\
\midrule
\multirow{4}{=}{\textbf{Preventing the Creation of New Inequalities Through GenAI Teaching Practices}}
& Academic inequalities from shortcut use 
& Teacher efforts to foster critical engagement with GenAI. \\

& Cultural biases in GenAI outputs 
& Teacher strategies to build critical awareness of bias. \\

& Peer inequalities from GenAI misuse 
& Teacher efforts to set clear guidance and norms. \\

& Excessive reliance on GenAI for emotional support 
& Teacher efforts to emphasize human care and relationships. \\
\bottomrule
\end{tabular}
\label{fig:RQ1-summary}
\end{table*}

Our findings show that the K–12 teachers we studied in the United States, South Africa, and Taiwan continuously consider both what should be taught about GenAI and how it should be taught to students during their teaching practices, promoting educational equality. Following this paper's scope defined in \autoref{def_genai_ed}, this section focuses specifically on teachers’ practices for teaching students about GenAI within K–12 classrooms (\textit{GenAI Education}). We do not address how GenAI could support teachers’ overall teaching practices as a profession (\textit{GenAI in Education}).

When teaching students about GenAI, we found that teachers adopted varied approaches, shaped by their own AI fluency levels and subject areas. Sometimes, teachers directly taught what GenAI is and how it works. Other classes emphasized hands-on interaction, guiding students to experiment with GenAI tools. Still others blended instructions on GenAI use into regular subject teaching, such as language arts or science, so that students learned how GenAI might support subject-specific learning.

Across these varied approaches, teachers framed their instructions around two interconnected goals related to educational (in)equality. First, they taught students to use GenAI to mitigate existing inequalities (e.g., accessing GenAI generated materials when textbooks or other resources were scarce) so that disadvantaged learners would not fall further behind (see \autoref{All}). Second, they aimed to prevent GenAI from creating new inequalities by teaching responsible use, critical evaluation, cultural sensitivity, and the importance of maintaining human connection (see \autoref{Pre}). These dual orientations show how teachers viewed GenAI as a double-edged sword. On the one hand, it could open new opportunities for students who are often excluded from advanced technologies. On the other hand, if introduced without care, it could reinforce or even widen divides (Table \ref{fig:RQ1-summary}). 

\subsection{Using GenAI Education to Alleviate Pre-Existing Educational Inequalities} \label{All}

Teachers consistently described that GenAI education can be used to mitigate inequalities that had long predated GenAI’s arrival in classrooms. Such inequalities, including unequal access to knowledge and opportunities, had historically taken multiple forms: some students lacked basic awareness of emerging technologies (see \autoref{awareness}); others struggled with shortages of learning resources (see \autoref{learning}); many came to school with weaker digital literacy skills that excluded them from computing participation (see \autoref{literacy}); and marginalized learners, such as those with disabilities or without access to advanced courses, were often denied pathways to develop essential competencies (see \autoref{marg}).

Through their instructions on GenAI, educators sought to turn disadvantages into opportunities for inclusion. They introduced GenAI to raise learners’ awareness of new technologies, enable learners to compensate for resource scarcity, broaden digital participation through its accessibility, and teach personalized GenAI use for diverse learning needs. In this way, teachers positioned GenAI education not only as instructions about a novel technology, but as a means to help learners counteract unequal starting points and participate more equally in future AI-integrated societies.

\subsubsection{Addressing Consciousness Inequality: Using GenAI Education to Bridge Gaps in Awareness and Understanding of Technology.} \label{awareness}

Across all three regions, interviewees reported that many students had never encountered emerging terms such as GenAI education or AI literacy, or lacked awareness of their importance, which risked excluding them from future opportunities where AI would be essential, in the views of our sampled teachers. Teachers viewed GenAI education as a specific way to reduce this consciousness inequality by making new technology visible in students’ lives. As T6 explained, \textit{“GenAI is definitely the trend of the future… I really hope the kids can learn about it.”} By introducing GenAI in her classroom, she gave students from families with little exposure to technology a first step toward recognizing its importance. Similarly, U3, a U.S. teacher nearing retirement, highlighted how teaching about GenAI pushed students to think seriously about their futures on \textit{“what are vocations like there's certain things you really can't replace by AI.” }For both T6 and U3, teaching GenAI was about correcting unequal levels of technological awareness. Teachers used GenAI education as a direct intervention to close pre-existing gaps in students’ recognition of and engagement with emerging technology.

\subsubsection{Addressing Resource Inequality: Using GenAI Education to Compensate for Scarcity of Learning Materials.} \label{learning}

Teachers sought to teach students to use GenAI to address long-standing material inequalities in students’ everyday access to learning resources. 

South African teachers described GenAI as a stopgap solution to the chronic shortage of print and learning materials that had long disadvantaged their students. In schools where textbooks were scarce, students were left with only fragmented information, making it difficult to study independently or practice beyond the classroom. S1 reminded his students that they had to rely on whatever was available: \textit{“They have to make use of the GenAI resources available to them to their advantage. Because they don’t have much textbooks.”} For S1, GenAI education became a way to give his students access to examples, explanations, and practice materials that would otherwise be out of reach. 

S10 also explained how he encouraged students to use GenAI to fill the gaps in the government-issued life science textbook: \textit{“Because what we get in the textbook is the bare minimum.”} In his classroom, students use AI tools as an extension of textbooks to see connections, clarify concepts, and access richer examples than the limited printed pages provide. Thus, in contexts where students cannot afford supplementary workbooks or access internet-based learning platforms, these students can identify GenAI as the new alternative way to approximate the resources that wealthier peers might take for granted.

From the South African teachers’ perspective, even if students sometimes leaned heavily on AI outputs, this was still preferable to the past situation of having no material at all. As a rural South African teacher, S5 explained, students remembered AI-guided assignments and improved their exam responses compared to before. The alternative, for S5, was silence: without GenAI help, students were unable to practice because they lacked the texts and examples that others considered basic. By introducing GenAI, teachers gave disadvantaged learners at least a minimal baseline of access to knowledge, and GenAI instruction helped soften the sharp edge of resource inequality, ensuring that the absence of books or print materials no longer automatically excluded students from the learning process.

\subsubsection{Addressing Digital Literacy Inequality: Using GenAI’s Accessibility to Broaden Computing Participation.} \label{literacy}

Teachers also recognized that pre-existing inequalities in digital literacy meant that some students were excluded from computing education altogether. They found that GenAI’s natural language conversational interface offered a new way in, lowering the barriers for learners who lacked coding skills or prior technical experience.

U8, teaching in Mississippi, contrasted her students’ encounters with GenAI to her own early struggles with operating systems or software such as MS-DOS and Geocities. Today’s learners equipped with knowledge on GenAI use, she noted, can \textit{“create their own website in 4 and a half minutes with 5 pictures in a theme.”} For her, this shift illustrated how GenAI redefined digital literacy: students no longer needed specialized programming knowledge but could participate through natural language. This opened a computing pathway for children who might otherwise have been excluded from.

In Taiwan, T1 deliberately designed computing learning activities that relied on GenAI’s ease of interaction. For students who struggled with traditional computing tasks, she introduced image-generation assignments, asking them to produce Studio Ghibli–style four-panel comics or invent brand spokespersons. \textit{“The results were really fascinating,”} she recalled. \textit{“Everyone completed the four comics successfully using the Ghibli style.”} T1 carefully scaffolded the students by providing well-designed example GenAI prompts, circulating in class to encourage participation, and ensuring that all students had the opportunity to succeed.

By enabling K-12 students to interact through natural language with GenAI, teachers provided those previously left behind with opportunities to express themselves using cutting-edge technology. Such activities contribute to building student confidence and developing digital skills in ways that are both accessible and engaging.

\subsubsection{Personalizing and Opening Opportunities for Marginalized Students Through GenAI Education.} \label{marg}

Teachers highlighted how GenAI could personalize learning and open opportunities that were especially important for marginalized groups who had long been overlooked by traditional education structures. T4, a special education teacher in Taiwan who works with students with diverse learning disabilities (including ADHD, OCD, autism, dyslexia, and intellectual disabilities), explained that her students often struggle with reading comprehension and find it difficult to follow standard study routines. Traditionally, this meant they were left behind in learning how to set goals, plan effectively, and manage their own studies. These skills are considered essential for independence but are rarely made accessible to learners with disabilities. To address this inequality, she demonstrated to her students how to utilize GenAI for self-regulated planning. She guided the students with learning disabilities in her GenAI instruction classes to ask ChatGPT: \textit{“I have a big exam coming up. I only plan to study one hour a day, so could you tell me how much I can cover in that hour?”} For T4, the value of this GenAI education practice extended beyond a single exam. \textit{“This whole planning-oriented skill can be practiced within GenAI… I believe this is the kind of skill special education kids can carry with them for life, and that is what matters most for special education kids to fit into society.”} 

\subsection{Preventing GenAI from Creating New Inequalities Through Teaching Practice}\label{Pre}  

While teachers embraced GenAI as a tool to mitigate existing gaps, they were equally attentive to risks of new inequalities. They highlighted four main concerns and enacted practices to address them. First, GenAI could create academic divides between students who critically engage and those who use it as a shortcut; teachers responded by designing activities to foster verification and reflective use (see \autoref{chose}). Second, it could exacerbate peer inequalities through bullying or manipulation. Teachers countered this with disciplinary action and classroom dialogues to establish ethical boundaries (see \autoref{peer}). Third, GenAI outputs often reinforced cultural biases that marginalized local voices; teachers addressed this by cultivating students’ critical awareness of representation and identity (see \autoref{culture}). Finally, teachers cautioned that overreliance on AI might replace human connection, which creates emotional and relational inequalities; they emphasized instead the irreplaceable value of human care (see \autoref{care}).

\subsubsection{New Academic Inequalities from Shortcut Use, and Teachers’ Strategies for Cultivating Critical Engagement.} \label{chose}

At the academic level, teachers were particularly concerned that GenAI could create new divisions based on patterns of use. Some students treated GenAI as a supplement to deepen understanding, while others relied on it as a shortcut to complete assignments without genuine learning. This potentially leads to a new inequality: students who developed habits of critical thinking versus those who fell into passive dependence on AI-generated answers.  

To counteract this tendency, teachers designed classroom activities that encouraged verification and reflection. In South Africa, when teaching students how to use GenAI correctly, S4 structured discussions around comparing GenAI outputs with official government-prescribed textbooks: \textit{“In our class discussions, we have prescribed books that the government gives us, so we compare those textbooks’ answers to the ones generated by AI.”} By grounding AI responses against an authoritative baseline, students were trained to see them not as unquestionable truths but as provisional inputs to be evaluated and debated. Such practices cultivated critical engagement and helped prevent the emergence of shortcut-driven inequalities in learning.

\subsubsection{Emerging Peer Inequalities from Misuse of GenAI, and Teachers’ Efforts to Set Guidance.} \label{peer}

Beyond individual study habits, teachers emphasized that GenAI could create new peer inequalities among students if its misuse were left unchecked. Learners with greater digital skills or weaker ethical restraint could weaponize GenAI tools, humiliating or manipulating peers, while more vulnerable classmates lacked the means to defend themselves.

U10, a high school teacher in the United States, described a case where her students used AI to manipulate a peer’s image and circulate it for humiliation. Rather than leaving the situation to escalate, she worked with the school principal to ensure accountability: students and caregivers were brought together, the incident was formally addressed as bullying, and consequences were enforced. U10 stressed that such interventions were essential to protect victims and to make clear that GenAI misuse had real ethical boundaries.  

In Taiwan, T6 observed her students experimenting with deepfakes, some in ways that hinted at possible recruitment into exploitative activities. Alarmed by these behaviors, she decided not to simply ban the practice but to confront it directly in her teaching. T6 encouraged them to analyze concrete cases of scams and misinformation, and asked them to reflect on how curiosity could be directed toward productive applications rather than risky or unethical ones. As she put it, \textit{“Will these kids end up in scam syndicates in the future? They are actually quite smart.”} T6 worked to channel students’ technical curiosity into responsible exploration rather than leaving them vulnerable to harmful pathways.

By actively disciplining harmful behavior (U10) and foregrounding the risks in classroom dialogue (T6), teachers sought to prevent GenAI from amplifying peer inequalities. Their efforts showed how protective guidance could transform potential harms into opportunities for ethical education.

\subsubsection{Cultural Biases in GenAI Outputs as a New Inequality, and Teachers’ Strategies for Critical Awareness.} \label{culture}

Teachers emphasized that, left unchecked, GenAI outputs often reinforced cultural biases, which could create new inequalities by normalizing dominant perspectives while erasing or distorting local voices and identities.

T10 in Taiwan illustrated this challenge through a simple case of linguistic variation. She noticed that GenAI often defaulted to Mandarin Chinese terms common in mainland China, rather than the expressions used in Taiwan. To address this, she deliberately turned such discrepancies into class activities: she showed students examples, asked them to identify differences, and facilitated discussions on why the AI made these choices. \textit{“In mainland China they say Bo-Luo, in Taiwan we say Feng-Li… I use this to make students aware of local linguistic differences in AI outputs.”} By explicitly drawing attention to these differences, T10 encouraged her students to critically examine AI outputs, validate their own linguistic identity, and understand that cultural variation is a form of richness often ignored by current GenAI systems.

S8 in South Africa raised a more systemic concern about representation. While teaching science, she often asked students to use GenAI to learn examples of successful scientists in their fields. However, S8 observed that GenAI tools frequently reproduced U.S.-centric perspectives, defaulting to white, Western figures when asked to portray success, 
\textit{“They don’t really have a piece of everyone in there… the only successful people are white, and then black are not.”} Rather than ignoring these biases, she directly integrated them into her lessons as opportunities for critical reflection when teaching students how to use GenAI for science learning, such as knowing more about local achievements. 

These three cases from South Africa and Taiwan illustrate how the low-resource status of a language or culture can create substantial mismatches in teachers’ adoption of GenAI for classroom use. By foregrounding issues of cultural erasure and representation, they used GenAI education not only to expose new inequalities but also to cultivate critical awareness that helped students see themselves, their languages, and their communities as valid and worthy of representation.

\subsubsection{Relying on GenAI for Emotional Support Instead of Interpersonal Relationships as a New Inequality, and Teachers’ Emphasis on Human Care.} \label{care}

Some teachers emphasized that GenAI should never substitute for human-to-human care and connection, warning that the social and emotional dimensions of learning could not be outsourced to machines. They recognized a new potential inequality: while some students would continue to benefit from authentic human support networks, others might come to rely too heavily on AI emotional interactions, leaving them socially isolated and emotionally underserved.

U10 shared a particularly striking example. She explained that when she tried to manage students’ behavior, such as discouraging excessive phone use or frequent late-night outings, some students turned to ChatGPT for advice on how to negotiate with her. Through conversations with students, what surprised her even more was that students were not just using ChatGPT for tactical advice in dealing with teachers; they were also frequently seeking interpersonal relationship guidance and even emotional support. As a high school teacher, U10 described herself as \textit{“an at-school mom, at-school nurses, therapists}” someone whose responsibilities extended far beyond delivering lessons. For her, teaching also meant learning to interact respectfully with others and developing the emotional resilience needed to navigate adolescence. Seeing students outsource this interpersonal and affective learning to ChatGPT made her question how the role of teachers, as moral guides, caretakers, and everyday counselors, might be shifting in the presence of GenAI tools. She began to intentionally teach students to recognize the broader social dimensions of GenAI use.

S10 in South Africa was especially vocal about this risk, grounding his reflections in the philosophy of Ubuntu, a Southern African concept that values relationality and mutual care. He reminded his students: \textit{“We actually go by the slogan that a person is a person by another person… some reassurance that you can only get from a human being, not an AI.”} Beyond simply warning them, S10 incorporated these values into his classroom practice. He created regular opportunities when teaching GenAI, ensuring that students experienced the irreplaceable qualities of human connection. By embedding Ubuntu into his pedagogy, he sought to counterbalance the temptation of AI as a quick emotional substitute and to prevent students from seeing GenAI tools as a replacement for the relationships that sustain education and personal growth.

\section{Findings: Continuing Inequalities Challenges in GenAI Education Beyond Individual Efforts (RQ2)} \label{RQ2}

\begin{table*}[t]
\centering
\small
\caption{Continuing inequalities challenges in GenAI education beyond individual efforts (RQ2)}
\setlength{\tabcolsep}{8pt}
\renewcommand{\arraystretch}{1.1}
\begin{tabular}{p{0.29\linewidth} p{0.64\linewidth}}
\toprule
\textbf{Constraint Category} & \textbf{Description} \\
\midrule

\multirow{3}{=}{\textbf{Infrastructural Constraints on GenAI Education}}
& (1) Increased device availability without reliable connectivity.\\
& (2) Insufficient basic digital literacy among students with limited access to technology.\\
& (3) GenAI tool designs that introduce new forms of inequality.\\
\midrule

\multirow{2}{=}{\textbf{Insufficient Training and Curricular Guidance}}
& (1) Uneven access to high-quality professional development for GenAI instruction.\\
& (2) A widening gap between GenAI development and educators’ instructional preparedness.\\
\midrule

\multirow{2}{=}{\textbf{Restrictive Social Norms}}
& (1) Resistance shaped by broader social and political norms.\\
& (2) Traditional parental attitudes and culturally embedded beliefs.\\

\bottomrule
\end{tabular}
\label{fig:RQ2-summary}
\end{table*}

While \autoref{RQ1} highlights how teachers actively teach GenAI in their classrooms to promote equality, this section reveals the persistent barriers that limited the reach and sustainability of these teachers’ GenAI education efforts towards educational equality. These challenges were not the result of a lack of teacher commitment; indeed, many went to great depth to scaffold participation, localize tools, and bridge digital gaps. Yet, structural inequalities in AI infrastructure (see \autoref{infra}), professional training (see \autoref{training}), and prevailing social norms (see \autoref{norm}) repeatedly constrained what teachers could achieve on their own. 

\subsection{Infrastructural Constraints on GenAI Education: From Technical, Socio-Economic, and Design Perspectives} \label{infra}

Teachers consistently emphasized that infrastructural conditions posed foundational barriers to GenAI education equality. First, unstable internet connectivity left some students unable to access GenAI tools outside of school, even with devices in hand (see \autoref{device}). Second, limited and uneven access to technology produced long-term differences in students’ digital fluency, disadvantaging those from low-income households who lacked everyday opportunities to develop basic skills (see \autoref{tech}). Finally, teachers highlighted that the cultural and linguistic design of GenAI systems often reproduced inequalities (see \autoref{genaiculture}).

\subsubsection{One major challenge was the disconnection between device distribution and reliable connectivity.} \label{device} 
In the United States, U10, a high school teacher in an urban public school, noted that while her school provided each student with a school-issued Chromebook, many students still lacked reliable home Internet. U10 shared, \textit{“We are a one-to-one school, meaning everyone gets a school-issued computer, a little Chromebook, but, you know, not everyone has the internet at home.”} During COVID-19, the school district had temporarily ensured access for students without home connectivity, but such support was not permanent. After schools reopened, some students again found themselves without access to digital learning platforms outside of school. This challenge actually extended beyond students: during our interviews with 10 South African teachers, six interview sessions had to be rescheduled or canceled due to power or internet outages. As S1 explained, \textit{“In South Africa’s winter, power lines often break, causing entire areas to lose both electricity and internet access.”} These examples highlight how infrastructural gaps persist even in contexts with device availability, creating uneven opportunities for students to engage with GenAI-enhanced learning outside of school hours, a problem that no teacher can solve simply by adjusting classroom practice. 

\subsubsection{Basic digital fluency to use GenAI tools is still missed by students without technology access out of school.} \label{tech} While GenAI tools lower the bar for computing participation, U8 observed that students from low-income households often lacked even the most basic computer skills: \textit{“Skills like how to use a keyboard, how to navigate the internet, how to use Google to search. Things that students would naturally pick up on, [they] don’t have those skills because they don’t have regular access to such resources at their house.”} For many children, growing up without steady access to computers or reliable internet meant missing out on countless small, everyday opportunities to build digital fluency, e.g., typing homework assignments, searching for information online, or troubleshooting common problems. These small practices accumulate over the years into intuitive confidence with technology, which wealthier peers often take for granted.

\subsubsection{Teachers across regions also highlighted that the design of GenAI tools introduced inequalities of its own.} \label{genaiculture}Many noted that these GenAI systems often reflected cultural and linguistic biases that marginalized underrepresented groups. For instance, Taiwanese teachers such as T8 and T10 found that GenAI outputs frequently defaulted to mainland Chinese vocabulary, which risked erasing local linguistic identities. South African teachers, such as S8, observed that GenAI tended to reproduce Western or U.S.-centric perspectives, often portraying authority in ways that excluded Black or African experiences: \textit{“(I asked AI to) generate a picture of a South African lady, and then, I am a Black person, it gives me a white person. I'm like, okay, make it… A black person, and now the black person looks like a thief… they don't really have a piece of everyone in there, because they are sort of one-sided.”} Additionally, most GenAI tools nowadays provide only a universal interface with little attention to accessibility. Special education teachers like T9 frequently pointed out accessibility issues such as small font size, excessive steps, and limited or poorly designed color options that prevent special education students from entering the doorway of GenAI education: \textit{“The most important thing is the font size. For special education students, larger text is better, but some GenAI tools don’t allow adjustments, so the text becomes very small.”} These details are precisely what HCI researchers and GenAI designers should pay closer attention to.

\subsection{Insufficient Training and Curricular Guidance for GenAI Teaching} \label{training}

Beyond infrastructural gaps, many teachers encountered a professional barrier when integrating GenAI education into their classrooms: the lack of training and pedagogical guidance for K-12 GenAI education. While some schools offered professional development (PD) sessions, access to these opportunities was uneven, and many existing programs failed to provide practical, curriculum-aligned strategies for teaching GenAI. As GenAI technologies continue to evolve rapidly, teachers in our study described how the gap between tool development and educator preparedness has only widened, in ways that individual effort alone cannot bridge.

S8, a teacher in a well-resourced school in South Africa, described how regular PD sessions had become central to their teaching culture. These weekly PD sessions were led by a dedicated innovation team and focused heavily on emerging technologies. \textit{“A new idea within a week… that’s why they organize it weekly, and I feel like it’s very useful, because we always learn.”} Yet this rapid pace of change, on the other hand, revealed the limits of relying on PD sessions alone. While the constant introduction of new tools fostered a culture of learning, the pressure to adopt and implement the latest GenAI technologies left some teachers feeling disoriented and fatigued, particularly when clear pedagogical guidelines were lacking. As S8 implied, what teachers learned one week might already feel outdated by the next. 

Such support was even less available in under-resourced schools. Teachers often reported having few opportunities to learn how to integrate GenAI content meaningfully into their curriculum. T1, a relatively new teacher, expressed feeling uncertain and underprepared: \textit{“As a new educator, I don’t have much experience. I often worry that I don’t know how to incorporate AI into the classroom. There’s no example to follow, no standard process I can refer to.”} Without clear GenAI teaching frameworks, she feared that introducing GenAI to students could, to some extent, backfire. T2 echoed this concern: \textit{“We talked about the ambiguous beauty of AI, the way it can look so complete even when it’s actually not correct. That means [after teaching students GenAI], teachers have to constantly check, revise, and give new instructions [to the students].”} T2 explained that, even when he told students how to use GenAI for tasks like reading comprehension in class, the final step always had to return to the teacher: \textit{“In the end, I still have to check whether the AI’s answers match what I taught in reading class.”}

Therefore, even committed teachers who were willing to experiment and self-teach were constrained by the absence of systemic training about GenAI education. Teachers in well-funded schools risked being swept along by the pace of technological change, while those in under-resourced schools lacked even the starting points to teach GenAI confidently. In both cases, the problem exceeded what individual teachers could fix on their own. 

\subsection{Restrictive Social Norms towards GenAI Education} \label{norm}
Beyond infrastructural and pedagogical challenges, teachers also highlighted how restrictive social norms posed significant obstacles to GenAI education. Teachers described two prominent patterns: resistance from conservative groups that opposed AI on moral or ideological grounds (see \autoref{poli}), and hesitation from parents and communities who insisted on traditional principles of parenting and feared that AI would expose students to harm (see \autoref{parent}). 

\subsubsection{GenAI education encountered resistance rooted in broader social and political norms.} \label{poli}These forms of resistance were often manifested through opposition from conservative groups, cultural taboos, and fragmented institutional support for AI-driven educational innovation. Such social forces could significantly constrain the introduction of AI into classrooms, particularly when they shaped policy decisions, public discourse, or internal school dynamics.

T2, a Taiwan teacher who was one of the few actively promoting GenAI at his school, described how certain conservative religious and political groups publicly opposed what they referred to as \textit{“advanced teaching concepts.”} These groups, he noted, were not necessarily representative of actual parents or students. \textit{“Some of them claim to be part of parent associations, but in reality, their children already run companies. They're in their 70s, and they’ve created fake parent leadership groups just to influence education policy.”} These actors, ranging from fringe religious sects to ideologically motivated political organizations, positioned themselves as guardians of tradition and morality. They actively campaigned against the teaching of digital technologies, particularly AI, in classrooms. According to T2, their motivations were often rooted in preserving religious or ideological authority. \textit{“They try to protect their conservative values by creating associations and putting on a show… They’ll do anything to stop new technologies or forward-thinking ideas from entering education.”} 

\subsubsection{In addition to social barriers, teachers also observed how parental attitudes and cultural beliefs strongly shaped the slow adoption of GenAI education.} \label{parent}

These forms of resistance often stemmed from generational differences and deeply rooted traditions, making them harder to address through infrastructure improvements alone. Unlike organized political groups, parental resistance was more scattered, but it had an everyday influence on whether students had opportunities to explore GenAI. 

S7 explained that many parents in South Africa faced resistance to AI because they themselves had grown up without exposure to such technologies. \textit{“Now, the reasons for rejection is because those parents grew up like me, without AI—it was only recently introduced. And there is not enough teaching for parents on how to accept AI and use it safely. They just imagine the worst kinds of stories they hear: AI will take our jobs, robots will destroy us.”} This lack of familiarity led many parents to focus on the dangers of AI rather than its potential benefits, creating an atmosphere of suspicion around its educational use. 

Beyond fears of job loss or automation, cultural values also shaped decisions within households about whether even to allow internet or AI tools at home. Growing up in a rural area, S7 drew on his life experience, noting that while many families could afford Wi-Fi, they chose not to install it, believing that internet access might encourage harmful behaviors in children. In this sense, parental resistance was not merely about economics but about prioritizing traditional principles of parenting over technological advancement. S7 further emphasized that such fears were often reinforced by local cultural narratives. Just as earlier waves of technological change, such as the introduction of 5G networks, were met with rumors about cancer or mind control, GenAI was sometimes cast as a mysterious and dangerous force. \textit{“Parents say ‘GenAI is becoming more human than me, it’s going to overtake me, it’s going to control us’. Some even think AI might break families apart, teaching husbands or children the wrong values instead of those passed down by tradition.”} These anxieties, deeply intertwined with cultural worldviews and even beliefs in magic, made it difficult for communities to view GenAI as a tool for education rather than a threat to society. 

In sum, restrictive social norms, whether organized by conservative groups or embedded in household decisions, represent one of the most persistent and subtle barriers to GenAI education equality.

\section{Findings: Toward More Equal GenAI Education (RQ3)} \label{RQ3}
Sampled teachers in our study reflected deeply on how they could make their own GenAI teaching more inclusive with stakeholders’ further support. These reflections ranged from advocating for school-level infrastructures that could broaden access, to suggesting company- and government-level changes that would make classroom use more accessible. While these ideas extended across multiple levels of the education ecosystem, including schools (see \autoref{School-Level}), technology companies (see \autoref{Company-Level}), and governments (see \autoref{Government-Level}), a practical starting point was to improve individual teachers’ own GenAI teaching in the future (Table \ref{fig:RQ3-summary}).

\begin{table*}[t]
\centering
\small
\caption{Stakeholder-specific supports required to effectively integrate GenAI education in K--12 contexts (RQ3).}
\setlength{\tabcolsep}{6pt}
\renewcommand{\arraystretch}{1.1}
\begin{tabular}{@{}p{0.1\linewidth}p{0.87\linewidth}@{}}
\toprule
\textbf{Stakeholder} &
\textbf{Support Required} \\
\midrule
\textbf{School} &
\textbf{Establish Infrastructural and Organizational Support Within and Between Schools} \\
& (1) Provide equal infrastructural access to all students. \\
& (2) Create organizational support (innovation centers and professional development) to promote teachers’ GenAI practice. \\
& (3) Foster cross-school collaboration to redistribute cutting-edge resources among schools. \\
\midrule
\textbf{Company} &
\textbf{Create More Relatable and Accessible GenAI Education Experiences} \\
& (1) Localize GenAI products to reflect sensitive cultural topics and local norms. \\
& (2) Design accessible GenAI education experiences. \\
\midrule
\textbf{Government} &
\textbf{Civilize GenAI Education into Society} \\
& (1) Make GenAI education a universal competency for all citizens. \\
& (2) Co-create GenAI education with professionals across disciplines. \\
\bottomrule
\end{tabular}
\label{fig:RQ3-summary}
\end{table*}

\subsection{School-Level Envisions: Establish Infrastructural and Organizational Support Within and Between Schools} \label{School-Level}

Teachers emphasized that schools, as the immediate institutional environment where GenAI is introduced in classrooms, play a pivotal role in ensuring that the technology supports, rather than undermines, educational equality through GenAI education. They highlighted three domains where school-level interventions are crucial for building inclusive GenAI education: (1) providing infrastructural access to ensure that disadvantaged students are not excluded; (2) creating organizational structures, such as innovation centers and professional development opportunities, to sustain equal use; and (3) fostering cross-school collaboration to redistribute resources and reduce systemic divides.

Providing basic infrastructure support, such as computers, is crucial for promoting equality, as it gives disadvantaged students access to GenAI tools they may not have at home. As S6 explained, \textit{“Those learners who don't have access to laptops at home do have access to our computer classrooms, and they are always most welcome to use it after school hours.”} Such efforts ensure that motivated learners who, as S6 put it, \textit{“don’t want to be in the situation that they are in currently… they’re striving to get a good education and build a career for themselves,”} are not excluded from AI-supported learning opportunities. 

Beyond hardware and software infrastructure, organizational supports such as innovation centers and professional development (PD) are also essential for supporting teachers' GenAI education practice. S8 suggested that if all public schools in South Africa established an innovation center with equipment and other resources, whether backed by governments or some dedicated persons, students in low-resource contexts would be better equipped to participate in the GenAI era, regardless of the economic position of each school. Similarly, U8 highly praised professional development for bridging resource divides. U8, teaching in a Title I school serving predominantly low-income students, joined Code.org AI workshop and immediately applied its lessons to introduce digital citizenship and GenAI practices in her classroom. T4, a special education teacher from Taiwan, also described how PD sessions led by experienced peers enabled her to adapt AI strategies and recommend appropriate tools for her K-12 students with disabilities. 

Furthermore, while educational inequalities rooted in economic and cultural disparities have long persisted, some schools are beginning to experiment with ways to redistribute resources across schools and mitigate the Matthew Effect. For instance, S8, a teacher at an Independent Boarding School in South Africa, described how her school benefits from rich resources and how its outreach center supports teachers and students in less advantaged schools by sharing tools and training. She personally taught students from low-resource schools how to use AI, noting: \textit{“Those kids… If they get the knowledge, if they can get the right stuff, anything is possible. They will be able to catch up.”} These forms of cross-institutional collaboration highlight the potential of schools not only to advance their own students’ opportunities but also to act as community anchors for promoting GenAI education equality more broadly.

\subsection{Company-Level Envisions: Create More Relatable and Accessible GenAI Education Experiences} \label{Company-Level}

Across regions, we noticed nine South African teachers and eight Taiwanese teachers highlighted the lack of cultural sensitivity in current GenAI tools, which can reproduce structural inequalities by making AI appear foreign, inappropriate, or untrustworthy to local communities. This can distance and demotivate learners from learning GenAI knowledge. Teachers we interviewed emphasized that GenAI companies have a critical role to play in alleviating these barriers by localizing their GenAI products, including fine-tuning models for regional contexts and embedding cultural knowledge. 

As South African teacher S7 explained, GenAI systems should be able to recognize sensitive cultural topics and respond in ways that reflect local norms: \textit{“If one researches a taboo subject in South Africa, AI must be able to be culture sensitive… educate whoever is researching to say, ‘This is not allowed in this area, and here are the reasons.’ That will be more helpful in societal adoption. Once parents realize that what they value is what AI values, and what they don’t value it will also not respond, there’s no danger in letting kids learn it.”} S7 further suggested that technology companies establish regional branches or monitoring teams that adapt tools like ChatGPT or Grok to local expectations. Taiwanese teachers raised similar concerns, particularly for indigenous learners. T10, who taught Atayal\footnote{one Taiwanese indigenous group} children, observed that GenAI tools often misrepresented indigenous culture, requiring her to consult elders or community teachers to correct errors. She emphasized that better cultural representation would make indigenous children more willing to engage with GenAI: \textit{“If AI companies could collect and represent local culture better, indigenous children in Taiwan would be more willing to engage with AI.”} To make GenAI concepts relatable, she used everyday scenarios to explain them: \textit{“Atayal students know how to use a gun safely because their parents told them it cannot be pointed at others. In the same way, I explain to my students that GenAI is just a tool [similar to a gun]: it depends on how and when you use it.”} These perspectives highlight that cultural localization is not just a technical upgrade, but an equality intervention in GenAI education. By reflecting learners’ languages, traditions, and lived experiences, GenAI tools can dismantle structural barriers that have long excluded underrepresented communities, and GenAI education can become a more relatable and practical experience. At the same time, localization helps prevent new inequalities such as cultural erasure, misrepresentation, or mistrust of AI, making GenAI education more inclusive and sustainable.

Moreover, companies should account for the needs of disabled users when designing accessible GenAI education experiences. As T9 suggested, features such as fewer clicks, larger fonts, shorter paragraphs, more colorful text, or even eye-tracking support could significantly lower barriers for special education students to engage with GenAI systems. Improving accessibility in this way not only benefits special education learners but also advances broader HCI efforts toward more inclusive GenAI user and learning experiences.

\subsection{Government-Level Envisions: Civilize GenAI Education in the Society} \label{Government-Level}

Teachers across all regions emphasized that promoting educational equality requires GenAI education to be promoted as a universal competency for all citizens, even including adults, not only as next-generation schooling content. S7 highlighted that improving adults’ AI literacy can indirectly expand infrastructural access for K–12 learners. This observation is plausible for our sampled teachers, as in regions with higher adult AI literacy, parents tend to be more supportive of their children’s use of and education in GenAI. For example, in Hsinchu, where many parents work in the tech industries, teachers reported stronger parental encouragement for children to learn and use AI (e.g., \textit{``In fact, most of their family backgrounds are quite good. Basically, they are children of engineers, so their parents pay a lot of attention to their GenAI education"} (T4); \textit{``they [students from high AI literacy family] even got more exposure to AI than me."} (T9)). In contrast, S7 found that: \textit{``some families can afford to have Wi-Fi at their home. But they opt not to do that… Others fear ‘my kids would engage in graphic content very early, so I don’t want them to access the internet, so I’m not going to have Wi-Fi’."} S5, teaching in a conservative South African community, explained that many parents rejected AI outright but readily accepted WhatsApp as part of daily life. Thus, S5 encouraged her students to access Meta AI through WhatsApp and present it as a regular messaging service. This strategy reassured parents that it was for schoolwork, showing how local reliance on familiar apps created a pathway for GenAI use in contexts where GenAI itself was distrusted.

Civilization-level AI education is not only about educating adults, especially parents, about AI. It also means co-creating school-based GenAI education with professionals across disciplines. For example, U10 described how inviting graphic artists to her class exposed students to authentic creative processes, and reflected on how this could be extended to include GenAI applications. As she explained, \textit{“how much graphic art is now AI-generated… it would have been such a cool thing to just touch on with the kids and be like, this is why we like it, or this is why we hate it.” }She further raised questions about transparency in that GenAI class, wondering whether society should require digital products to disclose their AI origins, which are the issues she believed professionals could introduce to students more effectively. Such practices not only demystify AI, but also ensure that even students in under-resourced schools can see how AI is transforming diverse careers and industries, preventing awareness of AI-enabled futures from becoming a privilege reserved only for elite learners.

\section{Discussion}

This study contributes to a growing understanding of how K–12 teachers navigate the promises and perils of GenAI in educational equality through their everyday GenAI education teaching practices. Our findings across three regions (the United States, South Africa, and Taiwan) highlight a threefold dynamic. First, teachers actively used GenAI education as a mechanism to alleviate pre-existing inequalities and to prevent the emergence of new ones (see \autoref{RQ1}). Second, despite these efforts, their work was repeatedly constrained by infrastructural gaps, limited training, and restrictive social norms that lay beyond individual control (see \autoref{RQ2}). Third, teachers further envisioned concrete pathways for the future of GenAI education, outlining guidelines for schools, companies, and governments to support equal access and meaningful use better (see \autoref{RQ3}). 

In the following discussion section, first, we reconceptualize GenAI education as a teacher practice of negotiating equality (see \autoref{Discussion1}). Second, we distinguish GenAI education from earlier forms of computing education, highlighting how its natural language interface lowers barriers for diverse learners but also introduces new equality challenges that demand teacher mediation (see \autoref{d2}). Third, we extend beyond prior studies on individual adoption to demonstrate that teachers’ equality-oriented practices are deeply constrained by structural conditions of infrastructure, training, and social norms, underscoring the need to bridge micro-level pedagogical work with macro-level systemic inequalities (see \autoref{d3}). Finally, we synthesize these findings into a set of design tensions that articulate broader HCI contributions, and derive design and policy implications for supporting equality-oriented GenAI education at scale (see \autoref{d4}, \autoref{implications}).

\subsection{GenAI Education as a Teacher Practice of Negotiating Educational Equality} \label{Discussion1}

Our first contribution is to reconceptualize GenAI education as a global teacher practice of negotiating educational equality. Existing research on AI education often positions AI as a tool for learning efficiency, personalization, or assessment \cite{srinivasa2022harnessing,han2025enhancing,troussas2025novel}. In these accounts, teachers are frequently portrayed as adopters of technological innovations, responsible for integrating tools into their classrooms. Based on the previous literature, our study further highlights that teachers do not simply adopt GenAI into classrooms: they actively mobilize it as a pedagogical resource to mitigate long-standing inequalities while simultaneously guarding against the emergence of new divides.  

This framing extends existing work in two important ways. First, it shifts the analytic lens from viewing GenAI primarily as a technical instrument to seeing it as a socially situated practice. Prior studies have shown how AI can optimize classroom management or individualize learning content \cite{han2025enhancing,kaswan2024ai,xu2025ai}, but rarely have they examined how teachers use AI to address structural inequalities that predate the technology. Our findings show that teachers approached GenAI not only as an instructional aid but as a means to intervene in enduring inequalities, such as lack of resources, weak digital literacy, or exclusion of marginalized learners, and to design safeguards against new inequalities arising from shortcut use, misuse, cultural bias, or emotional overreliance.

Second, our conceptualization positions teachers as central agents of sociotechnical negotiation. While prior HCI and learning sciences literature emphasizes systemic reforms, policy initiatives, or product design as levers of educational equality \cite{hutson2022artificial,uttamchandani2018equity}, our study shows that equality is also forged through the micro-practices of teachers: the activities they design, the norms they reinforce, and the cultural framings they make explicit in their classrooms. Rather than viewing teachers as passive implementers of externally designed reforms, we argue that they act as mediators of technological possibility and as guardians of educational values, deciding when GenAI functions as an equalizer and when its risks must be curtailed. This builds on critical perspectives in CSCW and HCI that foreground human actors’ interpretive work in sociotechnical systems \cite{xiao2025let,cheon2023powerful}, extending them to the context of K–12 education by showing how teachers’ local negotiations shape the equality trajectory of AI tools.  

Theoretically, this contributes a novel perspective to AI and education research by foregrounding the dual orientation of teachers’ practices, which alleviates old inequalities and prevents new ones, as the defining condition of GenAI education. Our findings reveal that GenAI’s role is continuously negotiated through teachers’ situated practices. This dual orientation suggests that GenAI does not inherently follow a trajectory toward equality or inequality. Instead, its social role is contingent, relational, and value-laden, shaped by teachers’ day-to-day pedagogical strategies. By conceptualizing GenAI education as a teacher practice of negotiating equality, we emphasize the interpretive and normative work of educators in shaping equal futures with emerging technologies.

\subsection{From Traditional Computing Education to GenAI Education: New Potentials for Equalities} \label{d2}

A second contribution of this study is to distinguish GenAI education from earlier waves of computing initiatives. Prior work on computing education often emphasized technical skill acquisition, such as programming, computational thinking, or understanding algorithmic processes \cite{xiao2024exploring,selwyn2019should,stamper2024enhancing,belmar2022review}. These approaches positioned students primarily as future workers who needed to master technical competencies to remain competitive in the digital economy. By contrast, our findings show that GenAI education introduces a new paradigm: students can engage with advanced AI systems through natural language, image prompts, and conversational interaction, without requiring specialized coding knowledge or technical expertise. This shift dramatically lowers barriers to entry, enabling broader participation in AI-related learning \cite{gu2025ai}. Teachers highlighted how GenAI’s accessibility allowed students who previously struggled with computing tasks, such as those with weak digital literacy or learning disabilities, to actively create, question, and reflect through AI-supported activities.

At the same time, GenAI education introduces new challenges that differ from those in earlier computing education. Whereas traditional computer education often focused on the divide between those with and without access to devices or programming instruction \cite{sutton1991equity}, teachers in our study described how GenAI created divides based on usage patterns, cultural representation, and social norms. Some students treated GenAI as a shortcut rather than a tool for deeper engagement, or used it to harm peers through bullying or manipulation. Others encountered cultural biases embedded in GenAI outputs, such as the erasure of local languages or the reproduction of Western-centric narratives. These dynamics illustrate that the inequalities of GenAI education are not only about who has access, but also about how AI systems are designed and how students are guided to critically engage with them \cite{pei2025formally}. In this sense, GenAI education requires pedagogical practices that differ from those of earlier computing initiatives, with teachers playing a crucial role in guiding students to navigate both the opportunities and risks of AI.

Theoretically, this distinction builds on and extends the critical computing education literature. Orlikowski \cite{orlikowski1992duality} suggests that technologies are simultaneously shaped by and shaping social practices. Our findings demonstrate this duality in practice: teachers not only adopt GenAI tools but also reinterpret their role in classrooms to address equality concerns. For example, whereas HCI and CSCW research \cite{star1999ethnography,dye2019if} highlights how infrastructural conditions are often revealed only through breakdowns or points of exclusion, teachers in our study intentionally foregrounded GenAI’s infrastructural and cultural limits to students, transforming them into teachable moments. They positioned themselves as interpreters of GenAI, making its use a socially situated negotiation of educational equality rather than a purely technical adoption. This extends sociotechnical perspectives by showing that teachers not only respond to breakdowns, but actively curate and reinterpret GenAI’s limitations as part of equitable learning practice.

GenAI education, here we argued, differs from prior computing and AI literacy efforts in two key ways: (1) it opens new potentials for inclusion by lowering technical entry barriers and enabling more diverse learners to participate, as seen in this study and in other CHI studies work with disadvanged learners \cite{han2024teachers, atcheson2025d}; and (2) it introduces new inequalities tied to usage practices, cultural bias, and relational dynamics that require proactive teacher mediation to mitigate widening disparities in different learners’ future AI readiness \cite{long2020ai} and reduce exposure to AI harms \cite{harvey2025don}. These differences underscore the need to conceptualize GenAI education not as a continuation of earlier computing curricula, but as a distinct domain where the stakes of equality and social participation are more immediate and more complex.

\subsection{Structural Constraints on Equality-Oriented GenAI Education} \label{d3}

Another contribution of ours is to extend beyond prior studies that conceptualize teachers’ engagement and practice with educational technologies primarily in terms of individual adoption, attitudes, or self-efficacy \cite{saienko2020impact,getenet2024students,clipa2023teachers}. While this body of work has shed light on how teachers decide whether and how to teach new tools, our findings highlight that teachers’ equality-oriented practices with GenAI are deeply embedded in, and often constrained by, structural conditions of infrastructure, training, and social norms. Our contribution thus advances theory by bridging the micro-level focus on teachers’ everyday classroom practices with the macro-level analysis of systemic inequalities. 

Specifically, teachers’ efforts were constrained by three forms of systemic inequality. First, infrastructural conditions such as unreliable connectivity, unequal digital fluency, and the design biases of GenAI tools shaped what teachers could realistically achieve in classrooms. Second, the absence of systematic training and curricular guidance left teachers to navigate GenAI integration largely on their own. Teachers in our study repeatedly underscored that without clear frameworks, they risked either overburdening themselves or inadvertently encouraging misuse. Third, broader social norms, ranging from organized opposition by conservative groups to diffuse parental skepticism, further constrained GenAI education. These findings resonate with critical studies of educational technology that emphasize how technologies are entangled with cultural values and political contestation \cite{heath2024more,hall2013educational}. Teachers were not simply dealing with technical adoption, but also with societal fears about morality, automation, or cultural preservation. In these contexts, individual teachers’ efforts to promote equality often collided with community-level resistance that they could not resolve alone.  

These findings clarify why equality-oriented GenAI education cannot rely on teachers’ individual commitment alone. While teachers actively negotiate GenAI’s role in alleviating and preventing inequalities through everyday pedagogical practices, the scope and sustainability of this work are fundamentally shaped by forces beyond the classroom. The constraints identified in this study, ranging from infrastructural gaps and uneven training to contested social norms, systematically delimit what teachers can achieve through local negotiation. These structural conditions echo prior work showing that educational technologies do not automatically translate into equitable outcomes, but must be interpreted and appropriated within broader institutional and cultural contexts \cite{ackerman2000intellectual,orlikowski1992duality,cao2025ai}. This gap between teachers’ equality work and the structural conditions that enable or constrain it directly motivates our design tensions and implications below. Supporting equitable GenAI education therefore requires moving beyond individual initiative toward systemic scaffolding, in which GenAI systems are designed with equality as a core value and policies intervene to redistribute resources, build capacity, and cultivate public trust.

\subsection{Design Tensions in Equality-Oriented GenAI Education}
\label{d4}

Synthesizing findings across RQ1–RQ3, we identify a set of design tensions that structure equality-oriented GenAI education. Making these tensions explicit helps explain why teachers’ practices are both creative and constrained, and why equality-oriented GenAI education cannot be reduced to teachers' best practices alone.

\subsubsection{Lowering barriers versus deepening new forms of inequality.}  
GenAI’s natural language interface lowers technical entry barriers and enables broader participation, particularly for students with limited digital literacy or learning disabilities (RQ1). However, the same accessibility also introduces new inequalities based on usage patterns, such as shortcut reliance, uneven critical engagement, and differential exposure to AI harms (RQ1). Teachers were therefore caught in a tension between promoting access and guarding against superficial or harmful use. This tension highlights that inclusion through accessibility does not automatically translate into equitable learning outcomes, and that GenAI systems designed to be “easy to use” still require pedagogical scaffolding to avoid reproducing stratified forms of participation.  

From a broader HCI perspective, this tension challenges the common assumption that lowering interactional barriers is inherently equality-promoting \cite{kim2025systematic,xiao2025institutionalizing,harrington2019deconstructing}. Our findings suggest that accessibility-oriented design decisions can simultaneously redistribute participation and stratify outcomes, calling for HCI research to more closely examine how ease of use reshapes patterns of engagement, dependence, and harm across different user groups.

\subsubsection{Personalization versus standardization of educational values.}  
Teachers leveraged GenAI to personalize learning for marginalized students, enabling flexible pacing, tailored explanations, and individualized planning (RQ1). At the same time, they encountered standardized outputs that reflected dominant cultural, linguistic, or epistemic norms, marginalizing local identities and minority experiences (RQ1, RQ2). This tension placed teachers in the role of cultural interpreters who had to correct, contextualize, or resist AI-generated representations. From a design perspective, this reveals a fundamental challenge: personalization at the level of individual learners can coexist with homogenization at the level of cultural representation, unless localization and cultural sensitivity are treated as core design commitments.  

For HCI, this tension extends beyond education by foregrounding personalization as a multi-level phenomenon. While systems may successfully tailor content to individual users, they can simultaneously standardize values and representations at scale \cite{li2025actions,xiao2025bridging,schecter2025role}. This insight urges HCI researchers to examine not only who personalization serves, but also whose cultural and epistemic assumptions are embedded and amplified through personalized AI systems.

\subsubsection{Teacher mediation versus structural dependency.}  
Across contexts, teachers actively mediated GenAI use by setting norms, disciplining misuse, foregrounding bias, and emphasizing human care (RQ1). Yet their capacity to sustain this mediation was repeatedly constrained by infrastructural gaps, insufficient training, and restrictive social norms beyond their control (RQ2, RQ3). This tension exposes a misalignment between the moral and pedagogical responsibility placed on teachers and the limited structural support available to them. While GenAI education is often framed as a matter of teacher competence or innovation, our findings show that equality-oriented mediation depends on systemic conditions that cannot be addressed through individual effort alone.  

This tension contributes a general insight for HCI and CSCW research on human-in-the-loop systems. It highlights a recurring pattern in which human actors are positioned as ethical and relational safeguards for AI systems, while the infrastructural and organizational conditions required to support this work remain underdeveloped. Rather than treating human mediation as an unqualified design asset, HCI must interrogate when such mediation becomes an uneven redistribution of accountability that obscures structural responsibility.

\subsubsection{Human care versus machine-mediated support.}  
Finally, teachers grappled with the tension between GenAI as a supportive tool and its potential to displace human relationships, particularly in emotional and relational domains (RQ1). While some students used GenAI as a source of reassurance or guidance, teachers emphasized that emotional care, moral development, and social learning remain fundamentally human responsibilities. This tension underscores that GenAI education is not only about knowledge or skills, but also about preserving relational values that underpin equitable education. Designing GenAI systems for educational contexts therefore requires careful attention to where automation should stop and where human judgment and care must remain central.  

More broadly, this tension informs HCI debates on automation boundaries in socially consequential domains. Our findings suggest that affective and relational dimensions of interaction are central to equality and inclusion when AI systems enter contexts involving guidance and moral development \cite{fan2025user,henriksen2025tell,yu2025principles,fan2024minion}.

\subsection{Design and Policy Implications} \label{implications} 

The design tensions identified above point to a central implication: equality-oriented GenAI education cannot be achieved through K-12 teachers’ classroom practice alone, nor through isolated improvements to AI systems. Instead, addressing these tensions requires coordinated interventions at both the design and policy levels. In this section, we translate the tensions between access and inequality, personalization and standardization, teacher mediation and structural dependency, and human care and machine support into concrete implications for GenAI system design and educational policy.

\subsubsection{Implications for GenAI System Design.}
Several of the tensions identified in \autoref{d4}---particularly those between lowering barriers and deepening new inequalities, personalization and standardization of values, and human care versus machine-mediated support---directly implicate the design of GenAI systems.

Our findings suggest that GenAI systems must be designed with equality as a central value rather than an afterthought, as teachers repeatedly pointed out that biased outputs and inaccessible interfaces constrained their ability to provide inclusive learning experiences. Such negative experiences illustrate where current systems violate key Human--AI Interaction guidelines, pointing to design issues that GenAI product designers must address \cite{amershi2019guidelines}. More specifically, these pitfalls show that design cannot stop at technical efficiency or surface-level personalization, but must attend to how accessibility, affordability, and cultural relevance interact with educational values.

For example, affordable education-oriented versions of GenAI should allow meaningful experimentation and iteration, rather than restricting disadvantaged learners to minimal or heavily constrained use. Likewise, GenAI models should incorporate diverse training data and localization features that reflect local languages, cultural references, and marginalized identities, ensuring that learners in Taiwan, South Africa, or elsewhere do not feel erased by homogenizing defaults. This directly responds to the tension between personalization and standardization identified earlier, and echoes postcolonial HCI arguments that technologies traveling across contexts must be redesigned with local infrastructures, languages, and epistemologies in mind to avoid reinforcing global inequities \cite{irani2010postcolonial,ahmed2015residual,ferreira2024examining}.

Finally, addressing the tension between human care and machine-mediated support requires design to move closer to pedagogy. GenAI systems for educational contexts should embed features that scaffold verification, critical reflection, and responsible classroom use, helping teachers teach not only how to operate GenAI but how to engage with it ethically and critically \cite{xiao2025bridging}. Teachers’ negative experiences also highlight the need for stronger AI literacy among developers, managers, and decision-makers, so that potential harms, especially those affecting vulnerable learners, can be anticipated rather than treated as afterthoughts, as emphasized by prior CHI research \cite{xie2025exploring}.

\subsubsection{Implications for Educational Policy.}
At the same time, the tension between teacher mediation and structural dependency makes clear that many constraints shaping equality-oriented GenAI education cannot be resolved through design choices or individual teacher effort alone. These challenges call for policy interventions at government, community, and school levels.

Infrastructural investment is needed to ensure that device distribution is paired with reliable connectivity, particularly in rural and low-income communities, a persistent challenge highlighted in HCI and CSCW research on infrastructural inequality, especially in cultural or geographic minority contexts \cite{sultana2019parar,star1999ethnography}. Without such investment, the promise of lowered interactional barriers risks reproducing new forms of exclusion.

Teacher capacity likewise requires systemic support. Professional development and corresponding assessments \cite{zhang2026assess} must go beyond tool introduction toward sustained pedagogical frameworks for responsible GenAI integration, enabling teachers to manage the tensions between access, critical engagement, and ethical use over time \cite{ding2024enhancing}. Our findings also underscore the importance of community trust: parental skepticism, cultural taboos, and political resistance shaped students’ access to GenAI, suggesting the need for public education and participatory approaches to policy-making, consistent with HCI and CSCW work on community-centered design \cite{villari2021community,zytko2022participatory}.

Finally, equity must be addressed not only within individual schools but across schools with highly uneven resources. Interviewees’ success stories illustrate the importance of cross-school resource sharing and capacity building to avoid a Matthew Effect, in which already privileged schools accelerate their GenAI adoption while under-resourced schools fall further behind. Policy interventions that support collaboration, redistribution, and long-term capacity building are therefore essential for translating teachers’ equality-oriented visions into sustainable practice.

\section{Limitations and Future Work}

This study has several limitations that suggest directions for future research. First, our analysis, based on semi-structured interviews with teachers in the United States, South Africa, and Taiwan, offers cross-cultural insights but is not representative of all contexts. Future work should examine additional regions with different political economies and technological infrastructures to understand how GenAI education practices evolve across diverse sociocultural landscapes. Second, while our interviews foreground teachers’ voices, they overlook the perspectives of other key stakeholders, including students, parents, policymakers, and technology developers. Since GenAI education is shaped by multi-level structures, future research should include these groups to develop a more holistic understanding of how equality-oriented practices are negotiated within and beyond classrooms. Third, our study examined teachers’ accounts of current and envisioned practices rather than measuring student outcomes or long-term impacts. Future research should empirically test whether GenAI education influences learning trajectories, digital literacy, or attitudes toward AI.

\section{Conclusion}

GenAI is rapidly entering K–12 classrooms worldwide, raising urgent questions about whether it will bridge or deepen existing educational divides. Drawing on interviews with 30 teachers from the United States, South Africa, and Taiwan, we demonstrate that teachers utilize GenAI education to address persistent inequalities while also mitigating new forms of exclusion. Their efforts, however, are constrained by infrastructural, training, and cultural barriers. Sustaining equality-oriented practices will require stronger support from schools, technology companies, and governments. Our cross-national findings underscore the importance of global perspectives, and future research should build on this work through longitudinal and multi-stakeholder studies to trace how GenAI education evolves.

\bibliographystyle{ACM-Reference-Format}
\bibliography{references}

\end{document}